\documentclass[epj]{svjour}
\usepackage{graphics}
\begin{document}
\title{Phase transitions in quark matter and behaviour of physical observables in the vicinity of the critical end point}
\titlerunning{Phase transitions in quark matter...}
\author{P. Costa\inst{1,}\inst{2}, 
				C. A. de Sousa\inst{1},
				M. C. Ruivo\inst{1}
				\and Yu. L. Kalinovsky\inst{2,}\inst{3}
}                     
%
%
\institute{Departamento de F\'{\i}sica, Universidade de Coimbra,
P-3004-516 Coimbra, Portugal
\and
Laboratory of Information Technologies, Joint Institute for Nuclear Research, Dubna, Russia
\and 
Universit\'{e} de Li\`{e}ge, D\'{e}partment de Physique 
B5, Sart Tilman, B-4000, Li\`{e}ge 1, Belgium}
\authorrunning{P. Costa, \emph{et al.}}
\date{Received: date / Revised version: date}
%
\abstract{
We study the chiral phase transition at finite $T$ and $\mu_B$ within the framework of the SU(3) Nambu-Jona-Lasinio (NJL) model. 
The QCD critical end point (CEP) and the critical line at finite temperature and baryonic chemical potential are investigated: the study of physical quantities, such as the baryon number susceptibility near the CEP, will provide complementary information concerning the order of the phase transition. We also analyze the information provided by the study of the critical exponents around the CEP. 
} 
\maketitle
%

Understanding the behaviour of matter under extreme conditions is one of the most challenging issues in the physics of strong interactions.   
In the last years, major theoretical and experimental efforts have been dedicated to the  physics of relativistic heavy--ion collisions, looking for signatures of the quark gluon plasma (QGP).
A region of particular interest, that has been so far explored, is the $T-\mu_B$ phase boundary in order to try to explore different regions of the QCD phase diagram.
The existence of the CEP in QCD was suggested in the end of the eighties, and its properties have been studied since then (for a review see Ref. \cite{Stephanov:2004}).
The most recent lattice results for the study of dynamical QCD with $N_f=2+1$ staggered quarks of physical masses indicate the location of the CEP at $T_E=162\pm2\mbox{MeV},\,\mu_E=360\pm40\mbox{MeV}$ \cite{Fodor:2004JHEP}, however its exact location is not yet known (the location of the CEP depends strongly of the mass of the strange quark). At the CEP the phase transition is of second order, belonging to the three-dimensional Ising universality class, and this kind of phase transitions are characterized by long-wavelength fluctuations of the order parameter. 

As pointed out in \cite{Hatta:2003PRD}, the critical region around the CEP is not pointlike but has a very rich structure. This critical region is defined as the region where the mean field theory of phase transitions breaks down and nontrivial critical exponents emerge. The size of the critical region is important for future searches for the CEP in heavy ion-collisions.
Some studies have been done in the SU(2) sector \cite{Hatta:2003PRD,Schaefer:2006} but less attention has been given to the effects of the strange quark. In this paper we aim to investigate the chiral phase transition and the CEP in quark matter with strange quarks.

We perform our calculations in the framework of the  three--flavor  NJL   model, including the determinantal 't Hooft interaction that breaks the $U_A(1)$ symmetry, which has the  following Lagrangian: %
\begin{eqnarray} \label{lagr}
{\mathcal L} &=& \bar{q} \left( i \partial \cdot \gamma - \hat{m} \right) q
+ \frac{g_S}{2} \sum_{a=0}^{8}
\Bigl[ \left( \bar{q} \lambda^a q \right)^2+
\left( \bar{q} (i \gamma_5)\lambda^a q \right)^2
 \Bigr] \nonumber \\
&+& g_D \Bigl[ \mbox{det}\bigl[ \bar{q} (1+\gamma_5) q \bigr]
  +  \mbox{det}\bigl[ \bar{q} (1-\gamma_5) q \bigr]\Bigr] \, .
\end{eqnarray}
Here $q = (u,d,s)$ is the quark field with three flavors, $N_f=3$, and
three colors, $N_c=3$, $\hat{m}=\mbox{diag}(m_u,m_d,m_s)$ is the current 
quark mass matrix and $\lambda^a$ are the Gell--Mann matrices, 
a = $0,1,\ldots , 8$, ${ \lambda^0=\sqrt{\frac{2}{3}} \, {\bf I}}$. 
The model is fixed by the coupling constants $g_S, g_D$ in the Lagrangian (\ref{lagr}), the cutoff parameter $\Lambda$, which regularizes the divergent integrals, and the current quark masses $m_i$.
For our numerical calculations we use the parameter set: 
$m_u = m_d = 5.5$ MeV, $m_s = 140.7$ MeV, $g_S \Lambda^2 = 3.67$, $g_D \Lambda^5 = -12.36$ and $\Lambda = 602.3$ MeV, and we follow the methodology presented in detail in Refs. \cite{Costa:2003PRC,Costa:2002PLB}.

\begin{figure*}[t]
\resizebox{0.47\textwidth}{!}{%
  \includegraphics{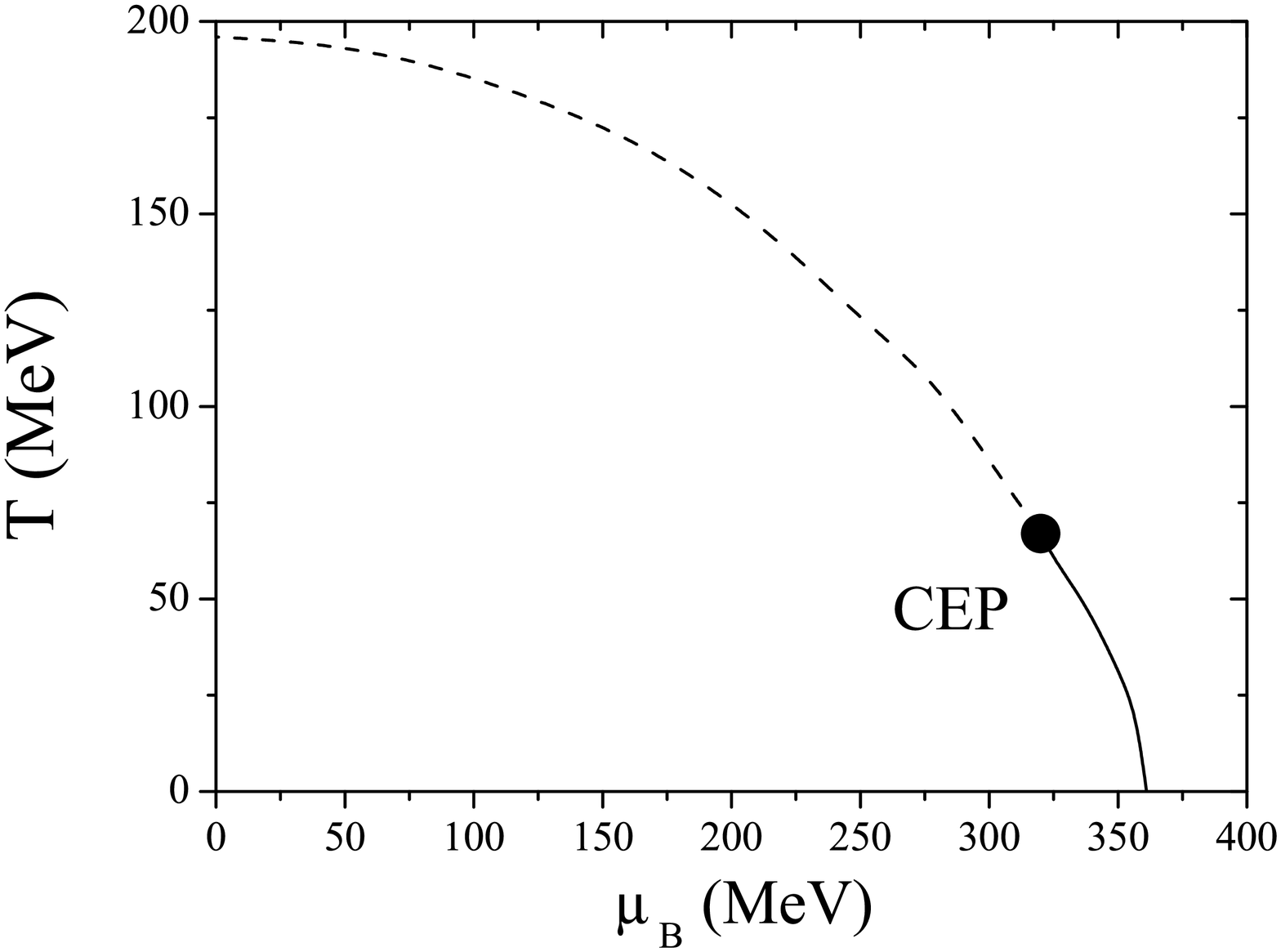}}
\resizebox{0.47\textwidth}{!}{%
  \includegraphics{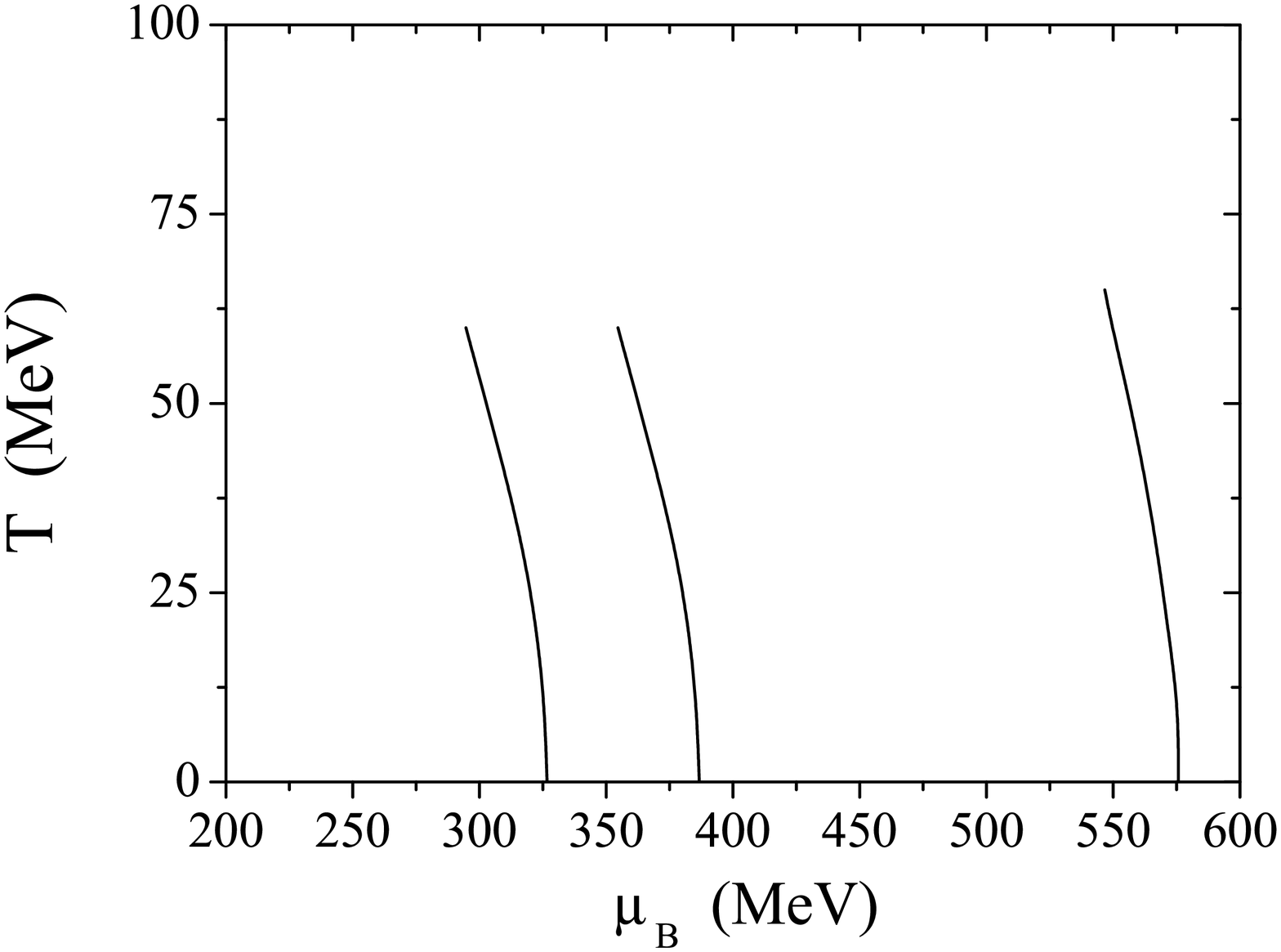}}
\caption{Phase diagrams in the $T-\mu_B$ plane. Left panel: $g_D=-12.36$. The solid line corresponds to first-order phase boundary and the dashed line corresponds to the crossover region. Right panel: $g_D=0$. The lines correspond to first-order phase boundaries of the $u$, $d$ and $s$ sectors (from left). We have used the accepted values $\mu_I=10$ MeV and $\mu_Y=12.5$ MeV.}
\label{fig:1}       
\end{figure*}

The baryonic thermodynamic potential has the following form:
\begin{equation}\label{tpot}
	\Omega (\mu_i ,T)= E- TS - \sum_{i=u,d,s} \mu _{i} N_{i}\,,
\end{equation}
where $E,\,S\, \mbox{and}\, N_i$ are, respectively, the internal energy, the entropy and the number of particles of the $i$th quark (the expressions are given in Sec. IV of Ref. \cite{Costa:2003PRC}).
The quark density for the $i$ quark is determined by the relation 
\begin{equation}
\rho_i = N_i/V = \frac{ N_c}{\pi^2}\int p^2 dp \left( n^-_i +n^+_i -1 \right)\theta(\Lambda^2-p^2)
\end{equation} 
where $n_i^{\pm}$ are the Fermi distribution functions of the positive (negative) energy state of the $i$th quark.
Along this work we impose the condition $\mu_e = 0$ so the chemical equilibrium condition is $\mu_u =\mu_d =\mu_s=\mu_B$ (with $\mu_B=(\mu_u+\mu_d+\mu_s)/3$).

In general, the baryon number susceptibility is the response of the baryon number density $\rho_B(T, \mu_i)$ to an infinitesimal variation of the quark chemical potential $\mu_i$:
\begin{equation} \label{chi}
	\chi_B (\mu_B, T) = \frac{1}{3}\sum_{i=u,d,s}\left(\frac{\partial 
	\rho_i}{\partial\mu_i}\right)_{T}.
\end{equation}
This observable is one of the most relevant quantities in the context of possible signatures for chiral symmetry restoration in the hadron-quark transition and for transition from hadronic matter to the QGP \cite{Stephanov:1998PRL}.

The nature of the chiral phase transition in NJL type models at finite $T$ and/ or $\mu_B$ has been discussed by different authors \cite{Costa:2003PRC,Buballa:2004PR}. At zero density and finite temperature there is a smooth crossover, at nonzero densities, different situations can occur (for details see Ref. \cite{Costa:2003PRC}). 

The phase diagram for the NJL model is presented in Fig. \ref{fig:1}, left panel. For zero temperature the transition is of first order. As the temperature increases, the first order transition line persists up to a critical endpoint. In the CEP the chiral transition becomes of second order. 

We will start our study by investigating the influence of the 't Hooft determinant in the $T-\mu_B$ plane in the SU(3) NJL model. Once for heavy-ion collisions the isospin chemical potential ($\mu_I$) is supposed to be non-zero, the possible existence of more than one CEP in the phase diagram was proposed in \cite{Toublan:2003PLB}. On the other hand, to study whether a kaon condensate is formed, non-zero hypercharge chemical potential ($\mu_Y$) is also important. Having these facts in mind, we start by taking $g_D=0$ and, if we fix nonzero values of $\mu_I$ and $\mu_Y$, there are three first-order phase transitions at low $T$ and high $\mu_B$ and thus three CEP as we can see in Fig. \ref{fig:1}, right panel. 
However, in our specific case, where $g_D=-12.36$, there is only one first order transition line and, consequently, only one CEP. This is a consequence of the flavor-mixing effects that cannot be neglected in the discussion of the phase diagram \cite{Frank:2003PLB}.

A bound to the size of the critical region around the CEP can be found by calculating the baryon number susceptibility ($\chi_B$) and its critical behaviour.
If the critical region of the CEP is small, it is expected that most of the fluctuations associated with the CEP will come from the mean field region around the critical region \cite{Hatta:2003PRD}.

\begin{figure*}
\resizebox{0.5\textwidth}{!}{%
  \includegraphics{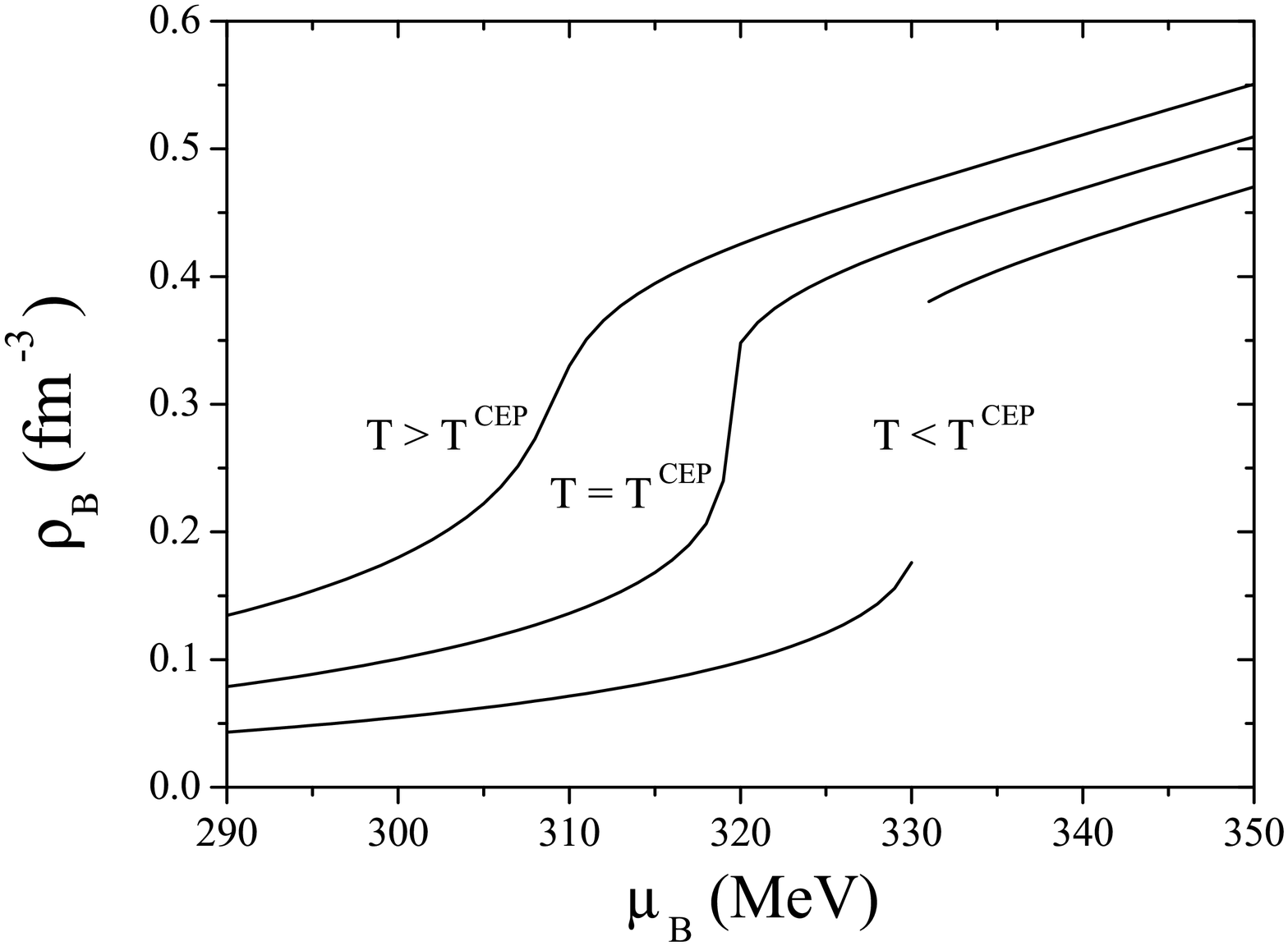}}
\resizebox{0.5\textwidth}{!}{%
  \includegraphics{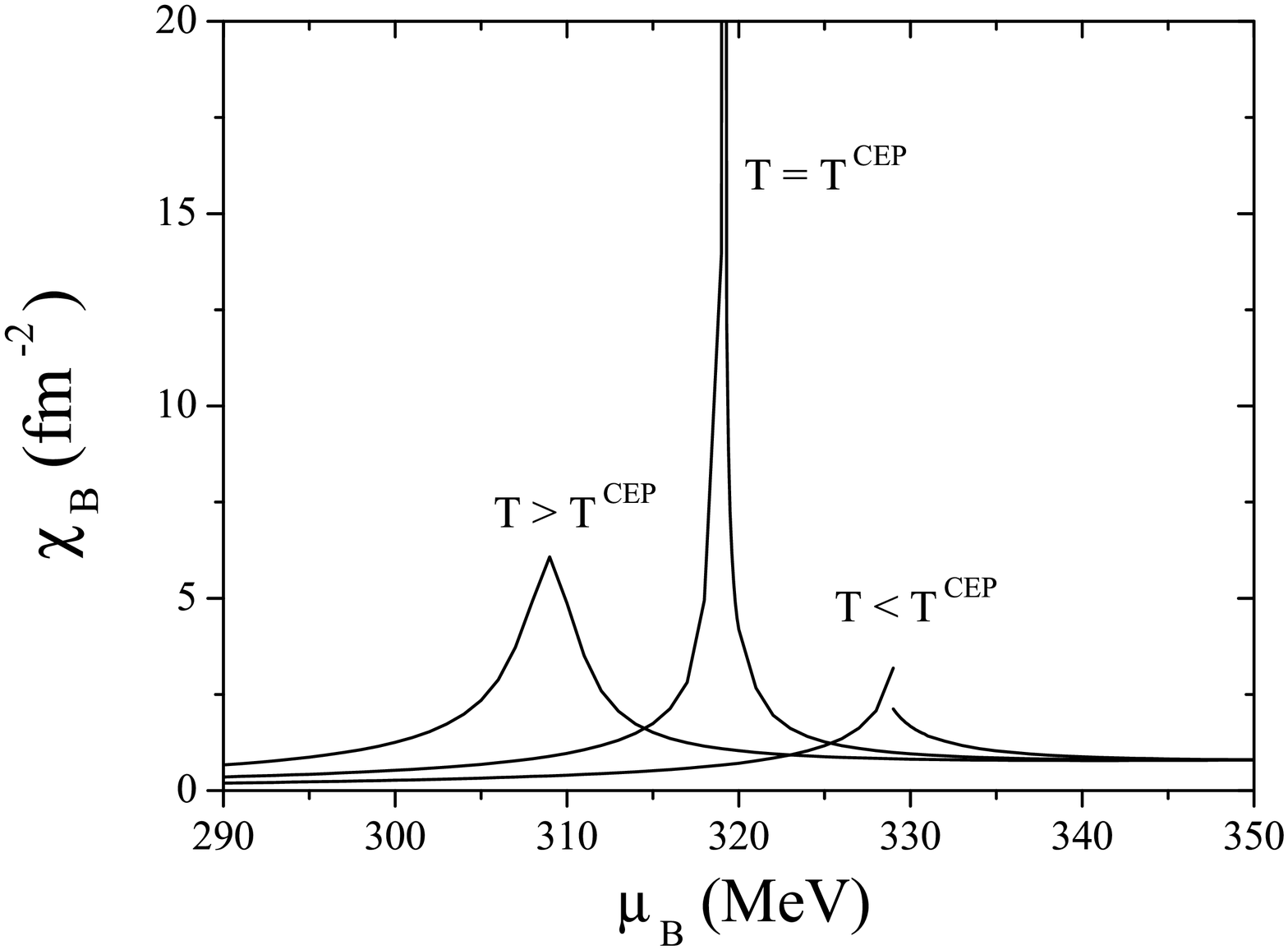}}
\caption{Baryonic density (left panel) and baryon number susceptibility (right panel) as a function of $\mu_B$ for different temperatures around the CEP: $T^{CEP}=67.7$ MeV and $T=T^{CEP}\pm10$ MeV.}
\label{fig:2}       
\end{figure*}

In the left panel of Fig. \ref{fig:2} the baryon number density for three different temperatures around the CEP is plotted. For temperatures below $T^{CEP}$ we have a first order phase transition and $\chi_B$ has a discontinuity (right panel of Fig. \ref{fig:2}). For $T = T^{CEP}$ the slope of the baryon number density tends to infinity at $\mu_B=\mu_B^{CEP}$ which implies a diverging susceptibility (this behaviour was found in \cite{Hatta:2003PRD,Schaefer:2006} using different models in the SU(2) sector). 
For temperatures above $T^{CEP}$, in the crossover region, the discontinuity of $\chi_B$ vanishes at the transition line, and the density changes gradually in a continuous way as we can see in the right panel of Fig. \ref{fig:2}. 

We will focus our attention on the critical behaviour of $\chi_B$ in the vicinity of the CEP. At the CEP the baryon number susceptibility diverges with a certain critical exponent. 
As pointed out in \cite{Griffiths:1970PR}, the form of the divergence depends on the route by which one approaches the critical point. If the path chosen is asymptotically parallel to the first order transition line, the divergence of the baryon number susceptibility scales with an exponent $\gamma_q$. 
In mean-field approximation, it is expected $\gamma_q=1$ for this path. For any other path not parallel to the first order line, the divergence scales with the exponent $\epsilon = 1-1/\delta$.
Once in the mean-field approximation $\delta=3$, we will have $\epsilon = 2/3$ and $\gamma_q > \epsilon$ is verified. The last condition is responsible for the elongation of the critical region, $\chi_B$ being enhanced in the direction parallel to the first order transition line; we verified that this is also true in the SU(3) NJL model \cite{Costa:2006}. 

\begin{figure}
\resizebox{0.5\textwidth}{!}{%
  \includegraphics{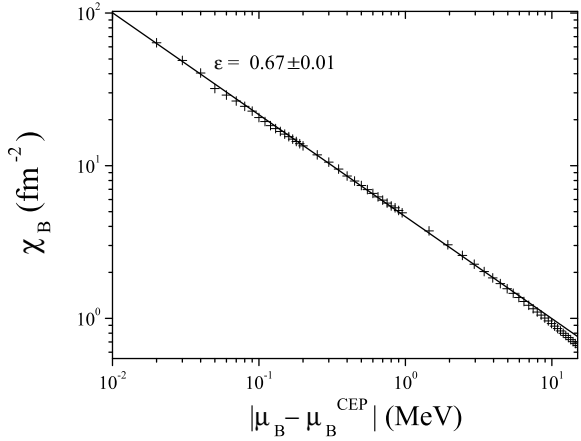}}
\caption{The baryon number susceptibility $\chi_B$ as a function of $|\mu_B-\mu_B^{CEP}|$ at fixed temperature $T^{CEP} = 67.7$ MeV.}
\label{fig:3}       
\end{figure}

For the baryon number susceptibility we will use a path parallel to the $\mu_B$-axis in the ($T,\mu_B$)-plane from lower $\mu_B$ towards the critical $\mu_B^{CEP}= 318.4$ MeV at fixed temperature $T^{CEP} = 67.7$ MeV. In Fig. \ref{fig:3} we plot $\chi_B$ as a function of $|\mu_B-\mu_B^{CEP}|$ close to the CEP. To calculate the critical exponent $\epsilon$ we will use a linear logarithmic fit 
\begin{equation} 
	\ln \chi_B = -\epsilon \ln |\mu_B -\mu_B^{CEP} | + const. ,
\end{equation} 
where the term $const.$ is independent of $\mu_B$. The result that we obtain is $\epsilon = 0.67\pm 0.01$, which is consistent with the mean field theory prediction: $\epsilon = 2/3$. 

We have analyzed the phase diagram in the SU(3) NJL which reproduces the essential features of QCD, such as a first order phase transition for low temperatures and the existence of the CEP. 
We verified that, in the chiral limit and differently from the SU(2) model, there is no TCP, which agrees with what it is expected: the chiral phase transition is of second order for $N_f = 2$ and first order for $N_f\geq3$ \cite{Pisarski:1984PRD}. A more rich scenario, in SU(3), is found when only the nonstrange quark masses are zero (the location of the CEP and the TCP depend strongly on the strange quark mass) \cite{Costa:2006}. Around the CEP we have studied the baryon number susceptibility which is related with event-by-event fluctuations of $\mu$ in heavy-ion collisions. 
We also conclude that in our model the critical exponent $\epsilon$ obtained is consistent with the mean field value $\epsilon=2/3$. 

\vspace{0.25cm}
Work supported by  grant RFBR 06-01-00228 (Yu. Kalinovsky) and by grant SFRH/BPD/23252/2005 from FCT (P. Costa).
\vspace{-0.3cm}


\end{document}